\documentclass[12pt]{article}

\usepackage{amsmath}
\usepackage{amsthm}
\usepackage{amssymb}
\usepackage{amscd}
\usepackage[british]{babel}
\usepackage{graphicx}
\usepackage{psfrag}
\usepackage{epsfig}
\usepackage{rotating}
\usepackage{times}

\theoremstyle{plain}

\newcommand{\boxend}{\flushright{$\Box$}}

\newcommand{\R}{{\mathbb R}}               % for real numbers
\newcommand{\C}{{\mathbb C}}               % for complex numbers
\renewcommand{\Re}{\mbox{\rm Re}}
\renewcommand{\Im}{\mbox{\rm Im}}

\newcommand{\w}{\omega}

\begin{document}

\begin{center}

{\bf \Large Dynamical Casimir Effect and the Black Body Spectrum}
\bigskip

%Partially-Reflecting Moving Mirrors and the Black Body Spectrum}

Jaume Haro$^{a,}$\footnote{E-mail: jaime.haro@upc.edu} and Emilio
Elizalde$^{b,}$\footnote{E-mail: elizalde@ieec.uab.es,
elizalde@math.mit.edu} \medskip

 $^a${\it \small Departament de Matem\`atica Aplicada I, Universitat
Polit\`ecnica de Catalunya \\ Diagonal 647, 08028 Barcelona, Spain}

$^b${\it \small Instituto de Ciencias del Espacio (CSIC) and
Institut d'Estudis Espacials de Catalunya (IEEC/CSIC)\\ Universitat
Aut\`{o}noma de Barcelona, Torre C5-Parell-2a planta, 08193 Bellaterra
(Barcelona) Spain}

\end{center}

\bigskip

\hfill  {\sl \small To Klaus Fredenhagen on the occasion of his 60th
Birthday:} {\it \small Herzlicher Gl\"{u}ckwunsch!}

\bigskip

\begin{abstract}
Creation of scalar massless particles in  two-dimensional Minkowski
space-time---as predicted by the dynamical Casimir effect---is
studied for the case of a semitransparent mirror initially at rest,
then accelerating for some finite time, along a specified
trajectory, and finally moving with constant velocity. When the
reflection and transmission coefficients are those in the model
proposed by Barton, Calogeracos, and Nicolaevici
[$r(w)=-i\alpha/(\w+i\alpha)$ and $s(w)=\w/(\w+i\alpha)$, with
$\alpha\geq 0$], the Bogoliubov coefficients on the back side of the
mirror can be computed exactly. This allows us to prove that,  when
$\alpha$ is very large (case of an ideal, perfectly reflecting
mirror) a thermal emission of scalar massless  particles obeying
Bose-Einstein statistics is radiated from the mirror (a black body
radiation), in accordance with previous results in the literature.
However, when $\alpha$ is finite (semitransparent mirror, a
physically realistic situation) the striking result is obtained that
the thermal emission of scalar massless particles obeys  Fermi-Dirac
statistics. Possible consequences of this result are envisaged.

\end{abstract}

%\vspace{1cm}

%{\bf Keywords.} Dynamical Casimir Effect, Partially Transmitting
%Moving Mirrors, Black Body Spectrum, Particle Creation,
%Bose-Einstein Statistics, Fermi-Dirac Statistics.

%\vspace{0.5cm}


%\maketitle
\bigskip

\noindent {\bf 1. Introduction.} The Davies-Fulling model
\cite{fd76,fd77} describes the creation of scalar massless particles
by a moving perfect mirror following a prescribed trajectory. This
phenomenon is also termed as the dynamical Casimir effect. Recently,
the authors of the present paper introduced a Hamiltonian
formulation in order to address some problems associated with the
physical description of this effect at any time while the mirror is
moving \cite{he06}; in particular, of the regularization procedure,
which turns out to be decisive for the correct derivation of
physically meaningful quantities. A basic difference with previous
results was that the motion force derived within the new approach
contains a reactive term---proportional to the mirror's
acceleration. This term is of the essence in order to obtain
particles with a positive energy all the time while the oscillation
of the mirror takes place, and always satisfying the energy
conservation law. Such result followed essentially from the
introduction of physically realistic conditions, e.g. a partially
transmitting mirror, which becomes transparent to very high
frequencies.

Here we will study a different aspect of the introduction of
physical, semitransparent mirrors, namely the particle spectrum
produced---in the conditions of the Fulling-Davies effect---by a
mirror of this sort which is initially at rest, then accelerates
during a large enough (but finite) time span, $u_0$, along the
trajectory defined in \cite{w85,cw87} (known to lead to a
Planck spectrum): \begin{eqnarray}
v=\frac{1}{k}(1-e^{-ku}) \label{f1}\end{eqnarray} (in light-like  coordinates,
where $k$ is some frequency), and finally, for $u\geq u_0$, is left
alone moving with constant velocity in an inertial trajectory.

Our interest will be to calculate the radiation emitted by the
mirror from its back (right) side. As is well-known,  a perfect
mirror that follows this kind of trajectory produces a thermal
emission of scalar massless particles obeying Bose-Einstein
statistics. More precisely, for $1\ll\w'/k\ll e^{ku_0}$  and
$1\ll\w'/\w\ll e^{ku_0}$ (with $\w'$ the frequency of an ingoing and
$\w$ of an outgoing  particle, respectively), the square of
the $\beta$-Bogoliubov coefficient satisfies \cite{c02b,n03,ha05}
\begin{eqnarray}
\left|\beta_{\omega,\omega'}^{R,R}\right|^2\equiv\left|
({\phi_{\omega,R}^{out}}^*;\phi_{\omega',R}^{in})\right|^2
\cong\frac{1}{2\pi\omega' k}
\left(e^{2\pi\omega/ k}-1\right)^{-1},
\end{eqnarray}
where this square of the $\beta$-Bogoliubov coefficient gives the average number of produced particles
in the $\w$ mode per unit of frequency. That is, the average number of produced particles
in the $\w$ mode, denoted ${\mathcal N}_\w$, is given by
 ${\mathcal N}_\w=\int_0^{\infty}d\w'\left|\beta_{\omega,\omega'}^{R,R}\right|^2$.

Here $\w'$ denotes the frequency of an ingoing particle (a particle coming from the past infinity), and
$\w$ the frequency of the outgoing one (particle going to the future infinity). Note that for the trajectory $(1)$
the ingoing mode suffers, after the scattering, a very high redshift, for this reason, in order to obtain $(2)$,
we need the above conditions (see for details \cite{ha05}, and Section (3a)).

Turning to the case of a partially reflecting mirror---in which we
will be mainly interested in this paper---in order to obtain the
radiation on the right hand side (rhs) of the mirror, we also need
to calculate the corresponding Bogoliubov coefficient, in this case:
$\beta_{\omega,\omega'}^{R,L}\equiv{({\phi_{\omega,R}^{out}}^*;\phi_{\omega',L}^{in})}^*$.

We thus first obtain the `in' modes on the rhs of the mirror when
the reflection and transmission coefficients are
$r(w)=\frac{-i\alpha}{\w+i\alpha}$ and $s(w)=\frac{\w}{\w+i\alpha}$,
with $\alpha\geq 0$, that is, when the Lagrangian density is given
by \cite{bc95,n01,c02a}
\begin{eqnarray}
{\mathcal L}=
\frac{1}{2}[(\partial_t\phi)^2-(\partial_z\phi)^2]-\alpha\sqrt{1-\dot{g}^2(t)}\phi^2\delta(z-g(t)),
\end{eqnarray}
where $z=g(t)$ is the trajectory in the $(t,z)$ coordinates.
\medskip

\noindent {\bf 2. Main results.} The main results of this paper,
some of them quite remarkable, are the following (for $1\ll\w'/k\ll
e^{ku_0}$ and $1\ll\w'/\w\ll e^{ku_0}$).
\begin{enumerate}\item In the perfectly reflecting  case, i.e.,
when $\w'\ll\alpha$, we obtain
\begin{eqnarray}\left|\beta_{\omega,\omega'}^{R,R}\right|^2\cong\frac{1}{2\pi\omega' k}
\left(e^{2\pi\omega/k}-1\right)^{-1}, \quad
\left|\beta_{\omega,\omega'}^{R,L}\right|^2\cong 0,
\end{eqnarray}
that is, a thermal radiation of massless particles obeying
 Bose-Einstein statistics arises.
\item In the perfectly transparent case, i.e.,when $\alpha\cong 0$, we have
\begin{eqnarray} |\beta_{\omega,\omega'}^{R,R}|^2 \cong 0,
\quad |\beta_{\omega,\omega'}^{R,L}|^2\cong 0.
\end{eqnarray}
In other words, there is no  particle production.
\item  In the physically more realistic case of a partially transmitting mirror
(transparent to high enough frequencies \cite{he06}), i.e., when
$\alpha\ll \w'$, what we obtain is
\begin{eqnarray} && \left|\beta_{\omega,\omega'}^{R,R}\right|^2\cong\frac{1}{2\pi\omega k}\left(\frac{\alpha}{\w'}\right)^2
\left(e^{2\pi\omega/ k}+1\right)^{-1}, \nonumber \\
&& \left|\beta_{\omega,\omega'}^{R,L}\right|^2\sim \frac{1}{\w\w'}
{\mathcal O}\left[ \left(\frac{\alpha}{\w'}\right)^2\right].
\end{eqnarray}
And, since
$\left|\beta_{\omega,\omega'}^{R,L}\right|\ll\left|\beta_{\omega,\omega'}^{R,R}\right|$,
we conclude quite surprisingly that a semitransparent mirror emits a
thermal radiation of scalar massless particles obeying Fermi-Dirac
statistics.
\end{enumerate}
Here it is important to emphasize that the word `statistics' refers to the $\beta$-Bogoliubov coefficient characterizing the spectrum of the radiated particles and not to the algebra obeyed by the creation and annihilation operators, that always satisfy the canonical anti-commutation relations. That is,
the original particles are bosons, but the spectrum of the radiated emission corresponds to fermionic ones. This could have some bearing on the local algebraic description of quantum fields \cite{11a}.

Given the novelty and potential importance of this result, we
thought we should devote the rest of the paper to provide a rigorous
and systematic proof of the same. Also, we will give hints to
possible interesting consequences and applications of our finding.
%e.g. to black hole physics (the Hawking radiation), at the end of this letter.
\medskip

\noindent {\bf 3. Proof of the results. (3a) Perfectly reflecting,
moving mir\-ror.} Consider a massless scalar field $\phi$  in
$2$-dimensional Minkowski space-time. Assume that the mirror
trajectory is ${\mathcal C}^1$ (once continuously
differentiable), and that it has the following form, in the
light-like coordinates $u\equiv t-z$ and $v\equiv t+z$,
\begin{eqnarray}
v=V(u)\equiv\left\{\begin{array}{ccc}
u,&\mbox{if}& u\leq 0,\\
%&&\\
\frac{1}{k}(1-e^{-ku}),&\mbox{if}& 0\leq u\leq u_0,\\
%&&\\
V(u_0)+A(u-u_0),&\mbox{if}& u\geq u_0,
\end{array}\right.\end{eqnarray}
with $A=e^{-ku_0}$. We also assume that $u_0\gg 1$. Note that this
trajectory can be written under the following form, too
\begin{eqnarray}
u=U(v)\equiv\left\{\begin{array}{ccc}
v,&\mbox{if}& v\leq 0,\\
%&&\\
-\frac{1}{k}\ln(1-kv),&\mbox{if}& 0\leq v\leq v_0,\\
%&&\\
U(v_0)+A^{-1}(v-v_0),&\mbox{if}& v\geq v_0.
\end{array}\right.\end{eqnarray}
For a perfectly reflecting mirror, the set of `in' and `out' mode
functions on the rhs of the mirror is \cite{dw75}
\begin{eqnarray} \begin{array}{c}
\phi_{\omega,R}^{in}(u,v)=\frac{1}{\sqrt{4\pi|\omega| }}
\left(e^{-i\omega v}-e^{-i\omega V(u)}\right)\theta(v-V(u)), \\
%\\
\phi_{\omega,L}^{out}(u,v)=\frac{1}{\sqrt{4\pi|\omega| }}
\left(e^{-i\omega u}-e^{-i\omega U(v)}\right)\theta(v-V(u)).
\end{array}
\end{eqnarray}
%and
%\begin{eqnarray}\left\{\begin{array}{c}
%\phi_{\omega,R}^{out}(u,v)=\frac{1}{\sqrt{4\pi|\omega| }}
%\left(e^{-i\omega u}-e^{-i\omega U(v)}\right)\theta(v-V(u)), \nonumber \\
%\phi_{\omega,L}^{out}(u,v)=\frac{1}{\sqrt{4\pi|\omega| }}
%\left(e^{-i\omega v}-e^{-i\omega V(u)}\right)\theta(u-U(v)).
%\end{array}\right.
%\end{eqnarray}
Our main aim now is to calculate the Bogoliubov beta coefficient
\begin{eqnarray}
\beta^{R,R}_{\w,\w'}\equiv
{({\phi_{\omega,R}^{out}}^*;\phi_{\omega',R}^{in})}^*, \ \ \  \w,\w'>0,
\end{eqnarray}
where the parenthesis on the rhs denotes the usual product
for scalar fields \cite{bd82}.

In order to compute this coefficient we choose
the right null future infinity domain ${\mathcal J}_R^+$;
since the trajectory is ${\mathcal C}^1$, we have
\begin{eqnarray}\label{ff}
\beta^{R,R}_{\w,\w'}&=&2i\int_{\R}du\
\phi_{\w,R}^{out}\partial_u\phi_{\w',R}^{in}=
\frac{1}{2\pi i\sqrt{\w\w'}}\frac{\w'}{\w+\w'} \nonumber \\
     &&-\frac{1}{2\pi
i\sqrt{\w\w'}}e^{-i\w u_0}e^{-i\w'V(u_0)}\frac{\w'A}{\w+\w'A}\\&&
-\frac{1}{2\pi k} \sqrt{\w'/\w} \int_{0}^{1-A}ds\,
(1-s)^{i\w/k}e^{-is\w'/k}.\nonumber
\end{eqnarray}
If we do the approximation $1\ll \w'/k\ll A^{-1}$  and
$1\ll \w'/\w\ll A^{-1}$, we arrive at
\begin{eqnarray}\label{eq1}
\hspace*{-5mm} \beta^{R,R}_{\w,\w'}&\cong&  \frac{1}{2\pi i\sqrt{\w\w'}} \nonumber\\&&
      -\frac{1}{2\pi k} \sqrt{\w'/\w}
\int_{0}^{1-A} \hspace*{-4mm} ds(1-s)^{i\w/k}e^{-is\w'/k}.
\end{eqnarray}
To obtain an explicit expression for the second term on the rhs,
we consider the domain
\begin{eqnarray*} \hspace*{-1mm} D\equiv \{z\in \C | \Re z\in [0,1-A], \Im z \in
[-\epsilon,0],  k/\w'\ll \epsilon\ll 1 \}\end{eqnarray*} and, going
through the same steps as in \cite{ha05}, we easily obtain that
\begin{eqnarray}
\beta^{R,R}_{\w,\w'}\cong \frac{1}{2\pi i
\sqrt{\w\w'}}e^{-i\w'/k}\left(\frac{ik}{\w'}\right)^{i\w/k}
\Gamma\left(1+i\w/k\right).
\end{eqnarray}
Finally, using
$|\Gamma\left(1+i\w/k\right)|^2=\frac{\pi\w/k}{\sinh\left(\pi\w/k
 \right)}$ (see \cite{as72}), we get the announced result, for a
perfectly reflecting mirror, that
\begin{eqnarray}
\left|\beta^{R,R}_{\w,\w'}\right|^2\cong \frac{1}{2\pi \w'
k}\left(e^{2\pi\w/k}-1\right)^{-1}.
\end{eqnarray}
\medskip

\noindent {\bf (3b) Partially reflecting moving mirror.} First, we
search for the co-moving coordinates $(\tau, \rho)$, that is, the
coordinates for which the mirror is at rest, $\tau$ being the proper
time of the mirror, and we take $\rho$ such that its trajectory is
given by $\rho=0$. Introducing the light-like coordinates
$(\bar{u},\bar{v})$, defined as
\begin{eqnarray}
\bar{u}\equiv \tau-\rho,\quad \bar{v}\equiv \tau+\rho,
\end{eqnarray}
we will calculate the mirror's trajectory in the coordinates
$(\bar{u},\bar{v})$.  Along this trajectory, the length element obeys
the identity \cite{op01}
\begin{eqnarray}
d\tau^2=d\bar{u}^2=d\bar{v}^2=V'(u)du^2=U'(v)dv^2.
\end{eqnarray}
Then, an easy calculation yields the relations
\begin{eqnarray}
\bar{u}(u)\equiv\left\{\begin{array}{ccc} u,&\mbox{if}& u\leq 0,\\
%&&\\
\frac{2}{k}(1-e^{-k\frac{u}{2}}),&\mbox{if}& 0\leq u\leq u_0,\\
%&&\\
\bar{u}(u_0)+\sqrt{A}(u-u_0),&\mbox{if}& u\geq u_0,
\end{array}\right.\end{eqnarray}
and
\begin{eqnarray}
\bar{v}(v)\equiv\left\{\begin{array}{ccc}v,&\mbox{if}& v\leq 0,\\
%&&\\
\frac{2}{k}(1-\sqrt{1-kv})&\mbox{if},& 0\leq v\leq v_0,\\
%&&\\
\bar{v}(v_0)+A^{-\frac{1}{2}}(v-v_0),&\mbox{if}& v\geq v_0.
\end{array}\right.\end{eqnarray}

When the mirror is at rest, scattering is described by  the
$S$-matrix (see \cite{he06,jr91} for more details)
\begin{eqnarray}S(\w)=\left(\begin{array}{cc}
{s}(\w)&{r}(\w)e^{-2i\w L}\\
{r}(\w)e^{2i\w L}&{s}(\w)\end{array}\right),\end{eqnarray} where
$x=L$ is the position of the mirror. This $S$ matrix is taken to be
real in the temporal domain, causal, unitary, and the identity at
high frequencies \cite{he06}. Correspondingly, the `in' modes in the
coordinates $(\bar{u},\bar{v})$ are (see also \cite{bc95})
\begin{eqnarray}
g^{in}_{\w, R}(\bar{u},\bar{v})&=&
\frac{1}{\sqrt{4\pi|\omega| }}s(\w)e^{-i\omega \bar{v}}\theta(\bar{u}-\bar{v})
\nonumber \\
     && \hspace*{-8mm} +\frac{1}{\sqrt{4\pi|\omega| }}
\left(e^{-i\omega \bar{v}}+r(\w)e^{-i\omega \bar{u}}\right)\theta(\bar{v}-\bar{u}), \nonumber \\
g^{in}_{\w, L}(\bar{u},\bar{v})&=&\frac{1}{\sqrt{4\pi|\omega| }}
\left(e^{-i\omega \bar{u}}+r(\w)e^{-i\omega \bar{v}}\right)\theta(\bar{u}-\bar{v})\nonumber \\
     &&+
\frac{1}{\sqrt{4\pi|\omega| }}s(\w)e^{-i\omega \bar{u}}\theta(\bar{v}-\bar{u}).
\end{eqnarray}
Note that the `in' modes in the coordinates $(u,v)$, namely
$\phi^{in}$, are defined
in the right null past infinity domain ${\mathcal J}^-_R$ by
\begin{eqnarray}
\phi^{in}_{\w,R}=\frac{1}{\sqrt{4\pi|\omega| }}e^{-i\w v}, \quad
\phi^{in}_{\w,L}=0,
\end{eqnarray}
and in the left null past infinity domain ${\mathcal J}^-_L$ by
\begin{eqnarray}
\phi^{in}_{\w,R}=0, \quad
\phi^{in}_{\w,L}=\frac{1}{\sqrt{4\pi|\omega| }}e^{-i\w u}.
\end{eqnarray}
From this definition, it is clear that $\bar{g}^{in}_{\w,
k}(u,v)\equiv g^{in}_{\w, k}(\bar{u}(u),\bar{v}(v))$, with $k=R,L$,
are {\it not} such modes. However,  the modes $\bar{g}^{in}_{\w, k}$
constitute in fact an orthonormal  basis of the space of solutions
to our problem. Consequently, if we use the fact that
$\bar{g}^{in}_{-\w, k}=\bar{g}^{in
*}_{\w, k}$, we obtain the following relation
\begin{eqnarray}\label{a}\hspace*{-3mm}
\phi^{in}_{\w,k}=\int_{\R}d\w'\chi(\w')(\bar{g}^{in}_{\w',
k};\phi^{in}_{\w,k})\bar{g}^{in}_{\w', k},
\end{eqnarray}
with $\chi(\w')$  the sign function.
To be remarked is the fact that Eq.~(\ref{a}) is to be interpreted as follows:
\begin{eqnarray}\phi^{in}_{\w,k}=\lim_{\lambda\rightarrow\infty}\int_{\R}d\w'\chi(\w')(\bar{g}^{in}_{\w', k};\phi^{in}_{\w,k})\bar{g}^{in}_{\w', k}
F_{\lambda}(\w'),\end{eqnarray} where $F_{\lambda}(\w')$ is a frequency cut-off,
for instance $\frac{\lambda^2}{\lambda^2+(\w')^2}$.

To calculate explicitly the `in' modes, we choose the coefficients:
 \begin{eqnarray} r(w)=\frac{-i\alpha}{\w+i\alpha}, \qquad
 s(w)=\frac{\w}{\w+i\alpha}, \end{eqnarray}
with $\alpha\geq 0$. In this case, on the rhs of the mirror we
obtain
\begin{eqnarray}
\phi^{in}_{\w,R}(u,v)&=&\frac{1}{\sqrt{4\pi|\omega| }}e^{-i\omega v}+
\phi_{\w, R}^{refl}(u), \nonumber \\  \phi^{in}_{\w,L}(u,v)&=&\phi_{\w,
L}^{trans}(u),
\end{eqnarray}
where
\begin{eqnarray}
\phi_{\w, R}^{refl}(u)=\left\{
\begin{array}{l}
\frac{1}{\sqrt{4\pi|\omega| }}\frac{-i\alpha}{\w+i\alpha}e^{-i\w V(u)}, \ \ \ \ u\leq 0,\\
\\
 \frac{1}{\sqrt{4\pi|\omega| }}\frac{-i\alpha}{\w+i\alpha}e^{-\alpha\bar{u}(u)}
 -\frac{2\alpha}{k\sqrt{4\pi|\omega|
}}e^{-i\frac{\w}{k}}\int_0^{\frac{k}{2}\bar{u}(u)} \\
ds\, e^{\frac{i\w
}{k}\left(s+1-\frac{k}{2}\bar{u}(u)\right)^2}
e^{-\frac{2\alpha s}{k}}, \ \ \ \ 0\leq u\leq u_0,\\
\\
\frac{1}{\sqrt{4\pi|\omega| }}\frac{-i\alpha}{\w+i\alpha}e^{-\alpha\bar{u}(u)}
 -\frac{1}{\sqrt{4\pi|\omega|
}}\frac{i\alpha}{\sqrt{A}\w+i\alpha} \\
\times \left[ e^{-i\w V(u)}-e^{-i\w
V(u_0)}e^{-\alpha(\bar{u}(u)-\bar{u}(u_0))}
\right] \\
%& \\
-\frac{2\alpha}{k\sqrt{4\pi|\omega|
}}e^{-i\frac{\w}{k}}e^{-\alpha(\bar{u}(u)-\bar{u}(u_0))}
\int_0^{\frac{k}{2}\bar{u}(u_0)}  \\
ds\, e^{\frac{i\w }{k}\left(s+1-\frac{k}{2}\bar{u}(u_0)\right)^2}
e^{-\frac{2\alpha s}{k}}, \ \ \ \ u\geq u_0,
\end{array}\right.
\end{eqnarray}
and
\begin{eqnarray}
\hspace*{-5mm} \phi_{\w, L}^{trans}(u)=\left\{
\begin{array}{l}
\frac{1}{\sqrt{4\pi|\omega| }}\frac{\w}{\w+i\alpha}e^{-i\w V(u)},
\ \ \ \ u\leq 0\\
\\
\frac{1}{\sqrt{4\pi|\omega| }}e^{-i\w u}
 +\frac{1}{\sqrt{4\pi|\omega| }}\frac{-i\alpha}{\w+i\alpha}e^{-\alpha\bar{u}(u)}
 -\frac{2\alpha}{k\sqrt{4\pi|\omega|
}} \\ \int^{\frac{k}{2}\bar{u}(u)}_0\hspace*{-1mm}ds
(s+1-\frac{k}{2}\bar{u}(u) )^{2i\w/k} e^{-\frac{2\alpha s
}{k}}\hspace*{-2mm}, \,  0\leq u\leq u_0\\
\\
\frac{1}{\sqrt{4\pi|\omega| }}\frac{-i\alpha}{\w+i\alpha} e^{-\alpha\bar{u}(u)}
+\frac{1}{\sqrt{4\pi|\omega|}}\frac{e^{-i\w u_0}}{\w+i\alpha\sqrt{A}} \\ \left[ \w
e^{-i\frac{\w}{\sqrt{A}}(\bar{u}(u)-\bar{u}(u_0))}+i\alpha\sqrt{A}
e^{-\alpha(\bar{u}(u)-\bar{u}(u_0))}\right] \\
%& \\
-\frac{2\alpha}{k\sqrt{4\pi|\omega|
}}e^{-\alpha(\bar{u}(u)-\bar{u}(u_0))}
\int^{\frac{k}{2}\bar{u}(u_0)}_0ds \\ (s+1-\frac{k}{2}\bar{u}(u_0)
)^{2i\w/k} e^{-\frac{2\alpha s }{k}}, \ \ \ \ u\geq u_0.
\end{array}\right.
\end{eqnarray}
Note that (as already advanced)
in the case of perfect reflection, that is, when $\alpha\rightarrow \infty$, we get
\begin{eqnarray}
\phi_{\w, R}^{refl}(u)\rightarrow -\frac{1}{\sqrt{4\pi|\omega| }}e^{-i\w V(u)}, \ \ \
\phi_{\w, L}^{trans}(u)\rightarrow 0,
\end{eqnarray}
and when the mirror is transparent, i.e.  $\alpha\rightarrow 0$, we obtain
\begin{eqnarray}
\phi_{\w, R}^{refl}(u)\rightarrow 0, \ \ \
\phi_{\w, L}^{trans}(u)\rightarrow \frac{1}{\sqrt{4\pi|\omega| }}e^{-i\w u}.
\end{eqnarray}

We are interested in the particle production on the rhs of the
mirror, for this reason we must now obtain, for $\w,\w'>0$,
%\begin{eqnarray}
%\hspace*{-3mm} \beta^{R,R}_{\w,\w'}\equiv {({\phi_{\omega,R}^{out}}^*;\phi_{\omega',R}^{in})}^*, \ \
%\beta^{R,L}_{\w,\w'}\equiv {({\phi_{\omega,R}^{out}}^*;\phi_{\omega',L}^{in})}^*,
%\ \ \w,\w'>0.
%\end{eqnarray}
%In order to obtain these products, we choose the right null infinity domain ${\mathcal J}^+_R$.
%Then, we have
\begin{eqnarray}
 \hspace*{-2mm} \beta^{R,R}_{\w,\w'}= {({\phi_{\omega,R}^{out}}^*;\phi_{\omega',R}^{refl})}^*, \ \ \
\beta^{R,L}_{\w,\w'}= {({\phi_{\omega,R}^{out}}^*;\phi_{\omega',L}^{trans})}^*.
\end{eqnarray}
We start by calculating $\beta^{R,R}_{\w,\w'}$,
%=2i\int_{\R}du\phi_{\omega,R}^{out}\partial_u \phi_{\omega',R}^{refl}$,
with the result
\begin{eqnarray}\label{eq2}
\hspace*{-2mm} \beta^{R,R}_{\w,\w'}&\cong&
\frac{1}{2\pi\sqrt{\w\w'}}\frac{\alpha}{\w'+i\alpha}\left[1
-\frac{\alpha}{k} \hspace*{-1mm} \int_A^1 \hspace*{-2mm} dx\,
x^{i\w/k-\frac{1}{2}}e^{-\frac{2\alpha}{k}(1-\sqrt{x})}\right]\nonumber\\&&
+\frac{\alpha}{2\pi k i\sqrt{\w\w'}}e^{-i\w'/k} \int_A^1dx\,
x^{i\w/k-\frac{1}{2}}e^{i\frac{\w'}{k}x}\nonumber\\&& \times\left[1
-\frac{2\alpha}{k}
\int_0^{1-\sqrt{x}}ds\, e^{i\frac{\w'}{k}(s^2+2s\sqrt{x})}e^{-\frac{2\alpha
s}{k}}\right].
\end{eqnarray}
Now, provided that $\w'\ll \alpha$, then Eq.~(\ref{eq2}) turns into
Eq.~(\ref{eq1}). Consequently, we precisely obtain
the same behavior as for a perfectly reflecting mirror. However, in
the case $\alpha\ll \w'$, we observe that
\begin{eqnarray}
\hspace*{-4mm} \beta^{R,R}_{\w,\w'}\cong \frac{\alpha}{2\pi k
i\sqrt{\w\w'}}e^{-i\w'/k}\hspace*{-1mm}\left(i\frac{k}{\w'}\right)^{i\w/k+\frac{1}{2}}
\hspace*{-2mm}\Gamma\left(\frac{1}{2}+i\w/k\right),
\end{eqnarray}
and using the identity $|\Gamma\left(\frac{1}{2}+i\w/k\right)|^2=\pi
/\cosh\left(
 \pi\w/k\right)$ (cf.~\cite{as72}), we conclude that
\begin{eqnarray}\label{f}
\left|\beta^{R,R}_{\w,\w'}\right|^2&\cong& \frac{1}{2\pi
k\w}\left(\frac{\alpha}{\w'}\right)^2
\left(e^{2\pi\w/k}+1\right)^{-1}.
\end{eqnarray}

Finally, a simple but rather cumbersome calculation yields the
 result
\begin{eqnarray}
\left|\beta_{\omega,\omega'}^{R,L}\right|^2\cong 0, \ \ \
\w'\ll\alpha,
\end{eqnarray}
and
\begin{eqnarray}
\left|\beta_{\omega,\omega'}^{R,L}\right|^2\sim \frac{1}{\w\w'}
{\mathcal O}\left[ \left(\frac{\alpha}{\w'}\right)^2\right], \ \ \
\alpha\ll\w'.
\end{eqnarray}
Note that, in the case $\alpha\ll\w'$ we indeed obtain the nice
feature that  the number of created particles in the $\w$ mode,
together with the radiated energies, are both finite quantities when
$u_0\rightarrow \infty$, in perfect agreement with the conclusions
in \cite{n03}. More precisely, for a partially transmitting mirror
the number of produced particles in the $\w$ mode, namely ${\mathcal
N}_{\w}$, is approximately
 \begin{eqnarray} {\mathcal
N}_{\w} \cong
\int_0^{\infty}d\w'\left|\beta_{\w,\w'}^{R,R}\right|^2.\end{eqnarray}
In order to calculate this quantity, we split the domain
$[0,\infty)$ into two disjoints sets, $[0,k)$ and $[k,\infty)$. In
the second domain we can do the approximation (\ref{f}), and  we
obtain
\begin{eqnarray}\label{g}
\int_k^{\infty}d\w'\left|\beta_{\w,\w'}^{R,R}\right|^2\cong
\frac{1}{2\pi
\w}\left(\frac{\alpha}{k}\right)^2
\left(e^{2\pi\w/k}+1\right)^{-1}.
\end{eqnarray}
In the other domain, assuming that $k\ll 1$,   we have $\w'\ll 1$
and thus for incident waves of a very low frequency the mirror
behaves like a perfect reflector; for this reason we can use the
formula (\ref{ff}). Then, a simple calculation yields
\begin{eqnarray}
\int_0^{k}d\w'\left|\beta_{\w,\w'}^{R,R}\right|^2\sim
{\mathcal O}\left(\frac{ k^2}{\w(\w^2+k^2)}\right),
\end{eqnarray}
and, since $k\ll 1$, we conclude that the number of produced
particles in the $\w$ mode is approximately
\begin{eqnarray}
{\mathcal N}_\w\cong\frac{1}{2\pi \w}\left(\frac{\alpha}{k}\right)^2
\left(e^{2\pi\w/k}+1\right)^{-1},
\end{eqnarray}
and the radiated energy ${\mathcal
E}\equiv\int_0^{\infty}d\w\hbar\w{\mathcal N}_{\w}$ is, with good
approximation,
\begin{eqnarray}
{\mathcal E}\cong \frac{\hbar\alpha^2}{4\pi^2 k}\ln 2.
\end{eqnarray}
This completes the proof of all the statements above.

\medskip

\noindent {\bf 4. Final comments.} It is necessary to remark that
there is a crucial difference with the case $\w'\ll\alpha$, where
the number of radiated particles in the $\w$ mode diverges
logarithmically with $u_0\rightarrow \infty$. In this situation the
physically relevant quantity is the number of created particles in
the $\w$ mode per unit time. This dimensionless quantity is finite
and its value is given by \cite{n03,ha05}
\begin{eqnarray}
\lim_{u_0\rightarrow \infty}\frac{1}{u_0}{\mathcal N}_\w=\frac{1}{2\pi}\left(
e^{2\pi\w/k}-1\right)^{-1}.
\end{eqnarray}

A second point is that we have started an additional calculation for a
bidimensional fermionic model with massless particles \cite{he07}. We have found
that in this situation the reverse change of statistics  happens,
namely the Fermi-Dirac statistics for the completely reflecting
case turns into the Bose-Einstein statistics for the partially
reflecting, physical mirror.

To finish, note again the remarkable fact that the problem we
addressed here could  be solved {\it exactly}, thus
successfully completing a challenging program initiated by Barton,
Calogeracos, and Nicolaevici \cite{bc95,n01,c02a} about ten years
ago. As a consequence, the results we have obtained are absolutely
solid---they do not hang on a perturbative expansion or
approximation of any sort.

The physical reason for this surprising
change of statistics may be found in the fact that the form of the
spectrum is actually determined {\it not} through the statistics of
the field but rather by the specific trajectory of the mirror and by
its interaction with the radiation field. A related, albeit different,
example of a phenomenon of this sort occurs in the case of an
electric charge following the same trajectory (\ref{f1}).
When the radiation field has spin $1$, the radiation emitted by
the charge obeys Bose-Einstein statistics, but when a scalar
charge, and consequently an scalar radiation field, is considered,
the emitted radiation will
obey the Fermi-Dirac statistics \cite{nr95}.

Finally, we must point out for completeness that another situation
where a somehow related feature occurs (but maybe of a different kind) is when
measuring the spectrum of a scalar field using a DeWitt detector
\cite{dw75,bd82} which follows a uniformly accelerated world-line in
Minkowski space-time. In this case, when the dimension of the
space-time is even, the Bose-Einstein statistics is obtained, however when
this dimension is odd the reverse change of statistics, to the Fermi-Dirac
one, takes place \cite{t86,u86}.

\medskip

\noindent {\bf Acknowledgments.}  This work was supported in part by
MEC (Spain), projects MTM2005-07660-C02-01 and FIS2006-02842,
and by AGAUR (Gene\-ra\-litat de Ca\-ta\-lu\-nya),
grant 2007BE-1003 and contract 2005SGR-00790. The investigation is partly based
on work done while EE was on leave at the Department of Physics and
Astronomy, Dartmouth College, 6127 Wilder Laboratory, Hanover, NH
03755, USA.

\end{document}